\documentclass[twocolumn,showpacs,amsfonts,aps,prc,floatfix,superscriptaddress]{revtex4-1}
\usepackage[colorlinks=true, pdfstartview=FitV, linkcolor=red, citecolor=blue, urlcolor=blue]{hyperref}
\usepackage{amsmath}
\usepackage{bm}
\usepackage{graphicx}

\begin{document}

%Title of paper
\title{Medium-induced optical effects for prompt photons}

\author{Akihiko Monnai}
\email[]{amonnai@riken.jp}
%\homepage[]{Your web page}
%\thanks{}
\affiliation{RIKEN-BNL Research Center, Brookhaven National Laboratory, Upton, NY 11973, USA}
\date{\today}

\begin{abstract}
Electromagnetic aspects of a QCD matter have been a hot topic in recent years.
High-energy heavy-ion experiments revealed that flow harmonics of direct photons are not explained by most hydrodynamic models. In this work I discuss possible effects of refraction by the hot medium since it can work as a lens which converts its geometrical anisotropy into momentum anisotropy of photons. Then elliptic flow and higher-order harmonics of prompt photons are estimated numerically. The results indicate that they could be small but have phenomenologically non-trivial consequences. 
\end{abstract}

% insert suggested PACS numbers in braces on next line
% insert suggested keywords - APS authors don't need to do this
%\keywords{}

\maketitle

%%%%%%%%%%%%%%%%%%%%%%%%%%%%%%%%%%%%%%%%%%%%%%%%%%%%%%%%%%%%%%
\section{Introduction}
\label{sec1}
\vspace*{-2mm}
%%%%%%%%%%%%%%%%%%%%%%%%%%%%%%%%%%%%%%%%%%%%%%%%%%%%%%%%%%%%%%

Heavy-ion collisions at high energies are profuse in the phenomenology of quantum chromodynamics (QCD). Several experimental evidences indicate that hadrons are deconfined to the quark-gluon plasma (QGP) in the Relativistic Heavy Ion Collider (RHIC) and the Large Hadron Collider (LHC) \cite{RHIC:summary,LHC:v2}.
Azimuthal momentum anisotropy of particle yields has been quite informative on the collective properties of the subatomic QCD medium created in those collisions. It is quantitatively characterized by the harmonics of Fourier expansion of azimuthal particle spectra \cite{Ollitrault:1992bk, Poskanzer:1998yz}. Hadronic flow harmonics $v_n^h$, especially the elliptic flow $v_2^h$, is used to quantify that the system is strongly-coupled and fluid-like \cite{Kolb:2000fha} because it is experimentally found to be large compared with the corresponding azimuthal geometrical anisotropy of the overlapping region of the colliding nuclei. Consequently, the medium is considered to be opaque in terms of the strong interaction, which is also supported by the observations of jet quenching and heavy quark diffusion. 

On the other hand, direct photon flow harmonics $v_n^\gamma$ is sensitive to the nature of the QCD matter during its time evolution as well as to the initial time of the hydrodynamic stage because the medium is believed to be electromagnetically transparent. 
Here direct photons consist of prompt photons, which are created in the hard processes at the time of collision, and thermal photons, which originate from the inelastic processes in the medium. $v_n^\gamma$ had been speculated to be much smaller than $v_n^h$ in hydrodynamic models due to the contributions from earlier stages with smaller or, in the case of prompt photons, vanishing flow anisotropy. 

Recent experiments at RHIC and LHC, however, have revealed that the quantity is much larger than the previous expectations \cite{Adare:2011zr,Lohner:2012ct}. 
The source(s) of the enhancement of direct photon flow harmonics is(are) not well known theoretically. There are continuing debates on the mechanics behind the excess of direct photon $v_2^\gamma$ \cite{Chatterjee:2005de, Chatterjee:2008tp, Chatterjee:2011dw, Holopainen:2011pd, Chatterjee:2013naa, Chatterjee:2014nta, Liu:2009kta, vanHees:2011vb, Dion:2011pp, Basar:2012bp, Bzdak:2012fr, Goloviznin:2012dy, Hattori:2012je, Liu:2012ax, Shen:2013cca, Muller:2013ila,Linnyk:2013hta,Monnai:2014kqa,McLerran:2014hza,vanHees:2014ida}. The recent discovery of large direct photon triangular flow $v_3^\gamma$ at RHIC suggests that the large momentum anisotropy is at least partially due to the properties of the hot medium itself. Uncovering the mechanism behind this ``photon $v_n$ puzzle" would be a quite important step towards the full understating of the phenomenology of heavy-ion collisions.

Following the precision analyses of chromodynamic aspects of the hot medium, the importance of electromagnetic aspects have been recognized in recent years. The direct photon, as mentioned earlier, is an electromagnetic probe which retains information of the medium during its space-time evolution. The phenomenological consequences of the ultra-strong magnetic field created by the spectators are also studied extensively. On the other hand, intrinsic electromagnetic properties of the QCD medium without the presence of the strong magnetic field is not fully understood, such as in-medium corrections to the electric permittivity and the magnetic permeability, possibly because they are often assumed small at the QCD energy scale.

In this paper, the effects of refraction on photons due to the QCD medium is investigated. The medium is indicated to be electromagnetically transparent in experiments, but can have a non-unity refractive index \cite{Klimov:1982bv}. This would bend the photons traveling through it according to its spacial geometry. The hot medium would then work as an ``optical lens" and the converging/dispersing effect is expected to affect the flow harmonics $v_n^\gamma$ non-trivially. The effect is expected to be larger for the photons created earlier, \textit{e.g.}, prompt photons. The total number of direct photons is not modified unless the index becomes imaginary in this approach. Numerical estimations are performed for prompt photon flow harmonics $v_n^{\gamma,\mathrm{prompt}} (p_T)$, which are vanishing in most conventional formalisms, for demonstration.

In Sec.~\ref{sec2}, optics for the hot QCD medium in heavy-ion collisions is presented. In-medium refraction is introduced for photons. Numerical estimations of the flow harmonics of prompt photons due to the optical effects are presented in Sec.~\ref{sec3}. Sec.~\ref{sec4} is devoted for discussion and conclusions. The natural unit $c = \hbar = k_B = 1$ and the Minkowski metric $g^{\mu \nu} = \mathrm{diag}(+,-,-,-)$ are used in the paper.

%%%%%%%%%%%%%%%%%%%%%%%%%%%%%%%%%%%%%%%%%%%%%%%%%%%%%%%%%%%%%%%
\section{Optics in Heavy-Ion Collisions%Refraction of hot QCD medium
}
\label{sec2}
\vspace*{-2mm}
%%%%%%%%%%%%%%%%%%%%%%%%%%%%%%%%%%%%%%%%%%%%%%%%%%%%%%%%%%%%%%%

\begin{figure}[tbp]%[H]
\begin{center}
 \includegraphics[width=0.2\textwidth]{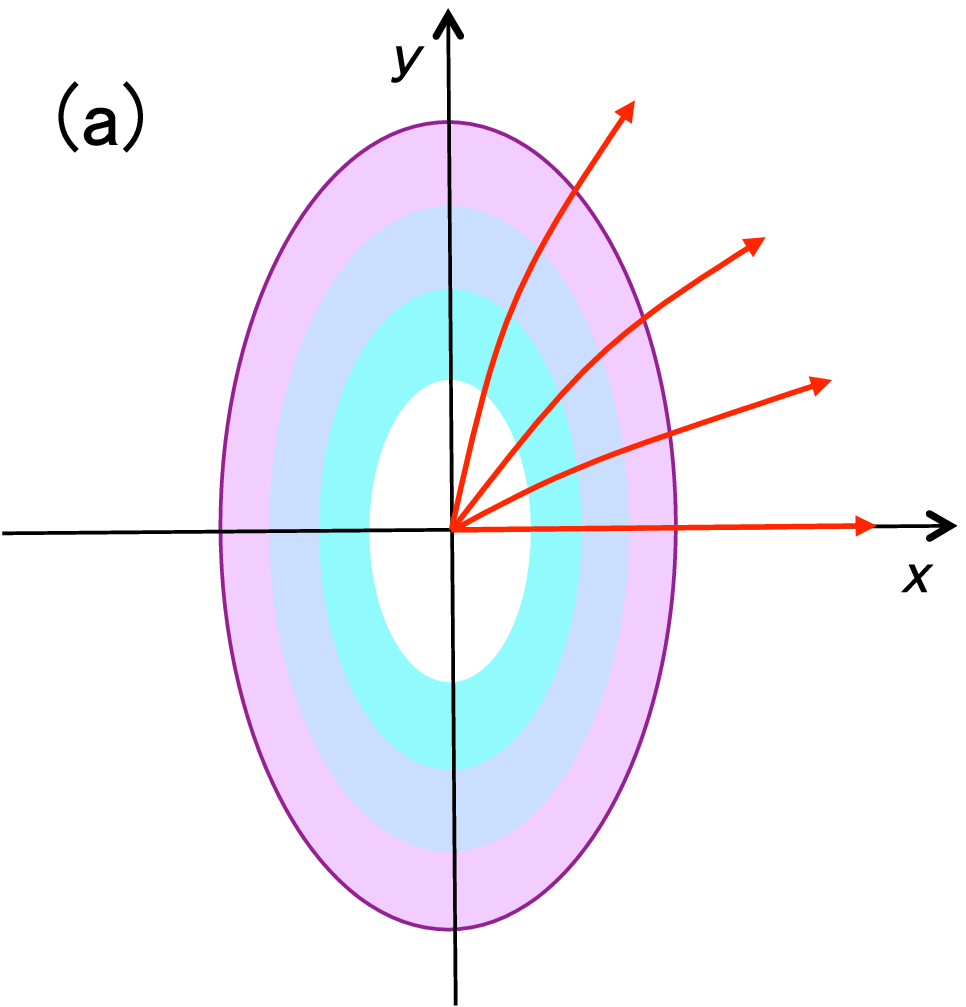}
 \includegraphics[width=0.2\textwidth]{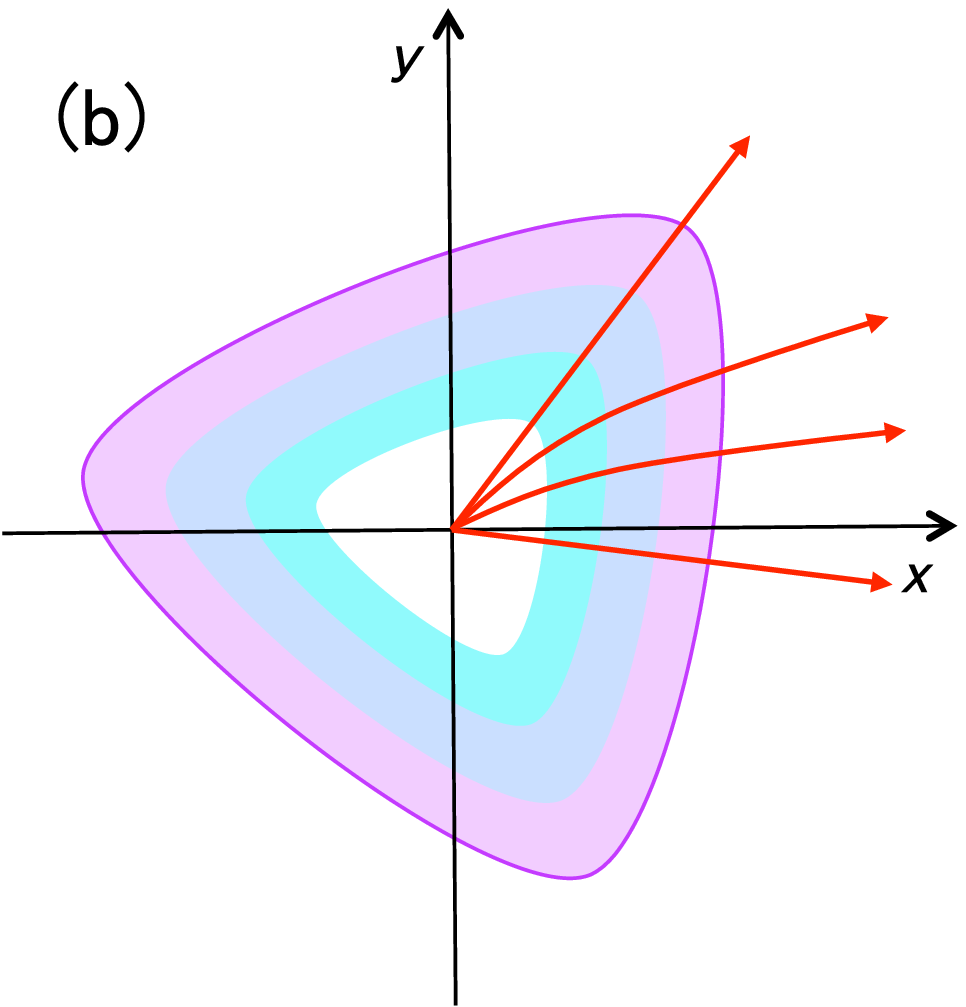}
\end{center}
\caption{(Color online) Schematic pictures of heavy-ion QCD media with (a) elliptic and (b) triangular geometries acting as lenses for the photons emitted from $x=y=0$. 
}
\label{Fig:1}
\end{figure}

A hot and dense medium is considered to be created in high-energy nucleus-nucleus collisions. I investigate the effects of medium refraction on the flow observables since it would directly correlate the spatial anisotropy of the medium and the momentum anisotropy of photons (Fig.~\ref{Fig:1}). Direct photons have to go through the medium and the refraction would be fully affected by the dynamical evolution of the medium because the typical radius of a heavy-ion nucleus ($R_\mathrm{Au} \sim 6.4$ fm and $R_\mathrm{Pb} \sim 6.7$ fm) is roughly the same as the typical lifetime of the hot medium ($\tau \sim 10$ fm/$c$) while the thermalization time is as short as $\tau_\mathrm{th} \sim 10^{-1}$-$1$ fm/$c$. 

%%%%%%%%%%%%%%%% Fig %%%%%%%%%%%%%%%%%%%%%
%%%%%%%%%%%%%%%%%%%%%%%%%%%%%%%%%%%%%%%%%%

Refraction in the medium with an inhomogeneous refractive index has been extensively studied in gradient-index optics. The path of a ray is given as
\begin{equation}
\frac{d^2 X}{d\tau ^2} = \frac{1}{2} \frac{dn^2}{dX} ,
\end{equation}
in Fermat's principle where $n$ is the refractive index and $X = (x,y,z)$ is the position on a ray. Here the refraction is considered at mid-spacetime rapidity, \textit{i.e.}, $z=0$.

The refractive index of a hot and dense QCD medium as a function of the temperature $T$ and the frequency $\omega$ is still not precisely known, though there are several prominent studies \cite{Klimov:1982bv,Liu:2011if,Jiang:2013nw}. Motivated by hard thermal loop calculations \cite{Liu:2011if} of the electrical permittivity $\varepsilon$ and the magnetic permeability $\mu$ for the QGP and the definition $n^2 = \varepsilon \mu$, here it is parametrized as:
\begin{eqnarray}
n^2 (T, \omega) = 1 - \frac{\omega_p^2(T)}{\omega^2} , \label{eq:index}
\end{eqnarray}
where $\omega_p$ is the characteristic or plasma frequency of the medium. The expression implies that the photons with higher momentum tend to be affected less by refraction, leading to a heavy-ion version of chromatic dispersion, \textit{i.e.}, the medium works as a prism. One can see that the medium is transparent only for $\omega > \omega_p$ and otherwise it is absorptive. The above expression is also valid for non-relativistic plasma. It should be noted, however, that this is a model and if a phenomenon such as pseudo-critical enhancement/reduction exists near the QCD crossover, the index would further deviate from that in vacuum. 

In general, it is hard to estimate $\omega_p (T)$. In the high $T$ limit of the QGP phase \cite{Liu:2011if}, one has roughly $\omega_p^2 \sim m_D^2$ where the Debye mass is $m_D^2 \sim e^2 T^2$ and the electromagnetic coupling is $e^2 = 4 \pi \alpha_\mathrm{EM}$. 
If the argument is na\"{i}vely extrapolated for $T \sim 300$ MeV, $\omega_p(T) \sim 10^{-1}$ GeV and the above expression would indicate that the medium is transparent for the frequency region of relevance in heavy-ion physics $\omega \sim 1$ GeV, which is consistent with the experimental implications. The medium absorbs photons below the momentum, which might lead to finite correction to the very-low momentum spectra. The latest RHIC experimental data of the photon $p_T$ spectra \cite{Adare:2008ab, Adare:2014fwh} pose a constraint $\omega_p < 0.5$ GeV on the model because no clear effect of absorption is found for the momentum above $0.5$ GeV. 

It should be noted that the photon frequency is modified in the flow $u^\mu = \gamma (1,\beta_x,\beta_y,0)$ in the medium by the relativistic Doppler effect as
\begin{eqnarray}
\omega = \frac{\omega_0}{\gamma (1 + \beta \cos \Delta \phi)} ,
\end{eqnarray}
at mid-rapidity where $\omega _0$ is the photon frequency without the effect, $\beta = (\beta_x^2 + \beta_y^2)^{1/2}$ is the transverse flow velocity and $\Delta \phi$ is the relative angle between the ray and the flow. This should modify the refractive index because Eq.~(\ref{eq:index}) is defined in the local rest frame.

It is also note-worthy that the phase velocity of the light is superluminal for the current model index, \textit{i.e.}, $v_\mathrm{ph} = 1/n > 1$ in the medium. This behavior is also found in Ref.~\cite{Klimov:1982bv}. The superluminality, of course, does not contradict causality because the group velocity $v_g = \partial \omega / \partial k = (1-\omega_p^2/\omega^2)^{1/2}$ is subluminal. The refractive index is related to the former while the propagation of information is associated with the latter. 

As mentioned earlier, the refractive index becomes imaginary and the direct photons can no longer propagate in the medium without attenuation below $\omega < \omega_p$. The exponential damping is characterized by the skin depth $\delta = 1/(\omega_p^2 - \omega^2)^{1/2}$. One has to be careful because $\delta_\mathrm{max} = 1/\omega_p$ can be comparable to the typical size of the medium in heavy-ion collisions, \textit{i.e.}, photons might be able to penetrate through the medium even if the attenuation occurs.

%%%%%%%%%%%%%%%%%%%%%%%%%%%%%%%%%%%%%%%%%%%%%%%%%%%%%%%%%%%%%%%%
\section{Numerical analyses}
\label{sec3}
\vspace*{-2mm}
%%%%%%%%%%%%%%%%%%%%%%%%%%%%%%%%%%%%%%%%%%%%%%%%%%%%%%%%%%%%%%%%

The procedure for numerical estimation is briefly summarized as follows. Prompt photons are produced according to the transverse coordinate dependence of the number of collisions right after the collision. Refraction for the in-medium photons becomes effective at the hydrodynamic initial time $\tau_\mathrm{th}$. 
The dynamical and inhomogeneous refractive index is calculated at each space-time point using a hydrodynamic model. The effect continues until the medium hits the hadronic freeze-out temperature, below which no refraction 
is assumed. Elliptic flow 
and higher-order flow harmonics of the prompt photons are then estimated. 
The effect of medium refraction on thermal photons will be discussed elsewhere.

The prompt photons are estimated with the scaled results of $p$-$p$ collisions. Here the parametrization \cite{Turbide:2003si} 
\begin{equation}
\omega_0 \frac{d^3 \sigma}{d^3 p} = 6495 \frac{\sqrt{s}}{(p_T)^5} \ \mathrm{pb/GeV}^2 ,
\end{equation}
for $x_T = 2p_T/\sqrt{s} < 0.1$ is employed. The actual emission rate as a function of a transverse position can be obtained by multiplying $(d N_\mathrm{coll} /dxdy)/ \sigma_{NN}^\mathrm{in}$ to the above expression where $N_\mathrm{coll}$ is the number of collisions and $\sigma_{NN}^\mathrm{in}$ is the inelastic cross section for nucleon-nucleon collisions. 
$N_\mathrm{coll}$ is given from the initial condition.
Experimental data indicates $\sigma_{NN}^\mathrm{in} = 42$ mb for $\sqrt{s_{NN}} = 200$ GeV \cite{Yao:2006px}. The validity of the above parametrization in a low momentum region should be considered carefully but would not affect the flow harmonics as the refractive index is not assumed to be a function of the intensity of light.

The space-time profile of the bulk medium or ``QGP lens" is given by the (2+1)-dimensional inviscid hydrodynamic model \cite{Monnai:2014kqa} with boost-invariance for the Au-Au collisions at $\sqrt{s_{NN}} = 200$ GeV. The equation of state is employed from the (2+1)-flavor lattice QCD calculation \cite{Borsanyi:2010cj}. The net baryon number density is assumed to be vanishing. The initial conditions are calculated with a newly-developed numerical code for Monte-Carlo Glauber model \cite{Miller:2007ri}. The energy distribution is constructed from the density distribution of the participant nucleons by setting the maximum energy density in the most central collision to 30 GeV/fm$^3$. The event-averaged smooth initial conditions are considered in the current analyses for numerical simplicity. It should be emphasized that they are employed for the purpose of demonstration. 
The medium effect after the initial time $\tau_\mathrm{th} = 0.4$ fm/$c$ and above the freeze-out temperature $T_f > 0.15$ GeV is taken into account.

It should be noted that na\"{i}ve installation of the refraction mechanism to a numerical hydrodynamic model is difficult, because the periodic lattice devision of a medium creates an artificial ``free path" for the photons traveling perpendicular to the lattice boundary. This increases the number of photons in specific directions, unphysically overestimating the flow harmonics with the same symmetry as the lattice regardless of its spacing. In this analysis, the temperature gradient field is smoothed with the bilinear interpolation method at each time-slice to avoid this behavior. This effectively makes the temperature field differentiable. The method is employed for calculating medium refraction, and space-time evolution, or energy-momentum conservation, is not affected.

Following the aforementioned hard thermal loop calculations, $\omega_p$ is treated as a parameter by introducing an auxiliary dimensionless factor $a$ defined in $\omega_p^2 = a^2T^2$. Since the initial medium temperature for RHIC is estimated as $T \sim 0.3$-$0.6$ GeV \cite{Adare:2008ab} and no obvious sign of absorption is found down to $p_T \sim 0.5$ GeV as mentioned earlier, $a < \mathcal{O}(1)$ is implied.

\subsection{Elliptic flow of prompt photons}

The azimuthal momentum anisotropy in the particle yield is quantified as a Fourier harmonics of a transverse momentum spectrum as
\begin{equation}
\label{eq:vngpt}
v_n^\gamma (p_T,y) = \frac{\int_0 ^{2\pi} d\phi_p \cos (n\phi_p - \Psi_n) \frac{dN^\gamma}{d\phi_p p_Tdp_T dy}}{\int_0 ^{2\pi} d\phi_p \frac{dN^\gamma}{d\phi_p p_Tdp_T dy}} ,
\end{equation}
in a differential form. Here $p_T$ is the transverse momentum, $y$ is the rapidity, $\phi_p$ is the angle in momentum space and $\Psi_n$ is the reaction plane angle. 

\begin{figure}[tb]
\includegraphics[width=3.0in]{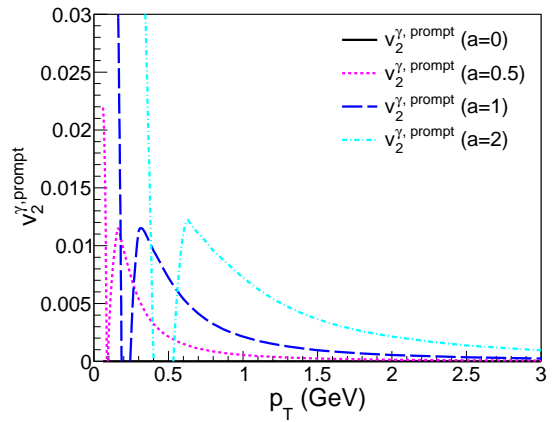}
\caption{(Color online) Prompt photon elliptic flow as a function of the transverse momentum due to refraction of a hot medium with $b = 6$ fm for $a = 0$ (solid line), 0.5 (dotted line), 1 (dashed line) and 2 (dash-dotted line). }
\label{fig:v2a}
\end{figure}

The elliptic flow of prompt photons at mid-rapidity is shown in Fig.~\ref{fig:v2a} for the parameters $a = 0$ (no refraction), $0.5, 1$ and $2$ for $b = 6$ fm. 
It can be found that the low-momentum photon elliptic flow above $\omega_p$ is created by the refraction of the medium, though the quantity is not large compared to that of experimentally observed direct photons. The effect is naturally more apparent for larger $a$. It should be noted, however, that prompt photon $v_2$ does not have to be as large as direct photon $v_2$ because there should be large contribution from thermal photons with both intrinsic and refractive anisotropy, and prompt photons might be playing only a supporting role. 

%%%%%%%%%%%%%%%%%%%%%%%%%%%%%%%%%%%%%%%%%%%%%%%%%%%%%%%%%%%%%%%
%%%%%%%%%%%%%%%%%%%%%%%%%%%%%%%%%%%%%%%%%%%%%%%%%%%%%%%%%%%%%%%

The ultra-low momentum region below $p_T \sim \omega_p$ has negative $v_2$ because the photons emitted inward are absorbed by a high-temperature core with $n(T, \omega)^2 < 0$. The dark ``core" is longer in the vertical direction so the effect of absorption becomes larger for the photons traveling horizontally. Once the emission source is within the core near $p_T \sim 0$, on the other hand, $v_2$ again becomes positive possibly because the photons emitted in the minor axis have more chance of penetrating through the medium than those in the major axis. Here the photon elliptic flow becomes negative because it is calculated using the hadronic reaction plane. In general it is not clear a priori whether the two reaction planes match.

\begin{figure}[tb]
\includegraphics[width=3.0in]{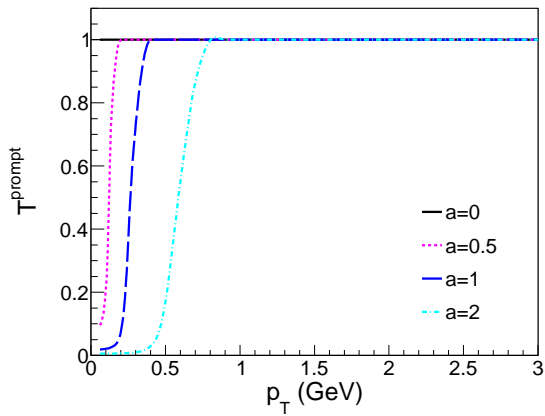}
\caption{(Color online) The ratio of prompt photon particle spectra with the medium refraction effect to the one without the effect as a function of the transverse momentum with $b = 6$ fm for $a = 0$ (solid line), 0.5 (dotted line), 1 (dashed line) and 2 (dash-dotted line). }
\label{fig:ratio}
\end{figure}

The transparency of the medium defined as 
\begin{equation}
\label{eq:vngpt}
\mathcal{T} = \bigg( \frac{dN_\mathrm{medium}^\gamma}{2\pi p_Tdp_T dy} \bigg) \bigg /\bigg( \frac{dN_\mathrm{vacuum}^\gamma}{2\pi p_Tdp_T dy} \bigg) ,
\end{equation}
is plotted for the prompt photons in Fig.~\ref{fig:ratio} to illustrate the region of applicability of the current approach. The absorption becomes effective near vanishing momentum region and it is stronger for larger $a$. As mentioned earlier, the experimental data imply transparency above $p_T \sim 0.5$ GeV, which prefers $a < 1$-$2$.

%%%%%%%%%%%%%%%%%%%%%%%%%%%%%%%%%%%%%%%%%%%%%%%%%%%%%%%%%%%%%%%
%%%%%%%%%%%%%%%%%%%%%%%%%%%%%%%%%%%%%%%%%%%%%%%%%%%%%%%%%%%%%%%

\begin{figure}[tb]
\includegraphics[width=3.0in]{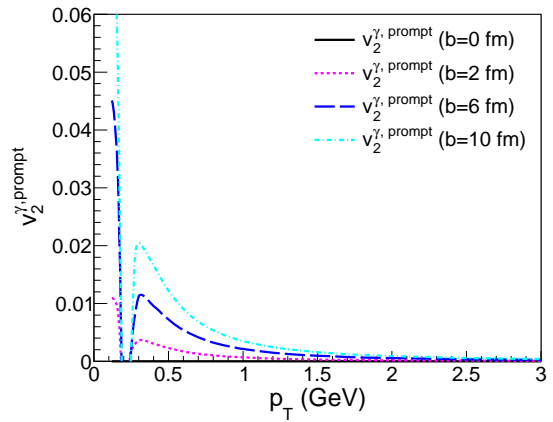}
\caption{(Color online) Prompt photon elliptic flow as a function of the transverse momentum due to refraction of a hot medium with $a = 1$ for $b = 0$ (solid line), 2 (dotted line), 6 (dashed line) and 10 fm (dash-dotted line). }
\label{fig:v2b}
\end{figure}

The eccentricity dependence of the prompt photon $v_2$ due to medium refraction is shown in Fig.~\ref{fig:v2b} for the impact parameters $b = 0, 2, 6$ and $10$ fm. The refraction parameter is set to $a = 1$. Corresponding initial geometrical anisotropies are $\varepsilon_2 = 0, 0.074, 0.202$ and $0.395$, respectively. Here the quantity is defined according to Ref.~\cite{Teaney:2010vd} as
\begin{eqnarray}
\varepsilon_n = - \frac{\langle r^n \cos[n(\phi - \Phi_n)] \rangle}{\langle r^n \rangle}, \label{eq:en}
\end{eqnarray}
for $n > 2$ where the bracket denotes spatial average. $r$ is the position of a participant nucleon and $\Phi_n$ is the minor orientation angle of the spatial geometry given by
\begin{eqnarray}
\Phi_n = \frac{1}{n} \arctan \frac{\langle r^n \cos(n \phi) \rangle}{\langle r^n \sin(n \phi) \rangle} + \frac{\pi}{n} ,
\end{eqnarray}
which is set to vanishing as the nucleons are rotated so that $\Phi_n = 0$ before taking average over events.
The numerical results imply that the optical $v_2$ becomes larger for more eccentric collisions, and the magnitude is roughly proportional to $\varepsilon_2$. 

%%%%%%%%%%%%%%%%%%%%%%%%%%%%%%%%%%%%%%%%%%%%%%%%%%%%%%%%%%%%%%%
%%%%%%%%%%%%%%%%%%%%%%%%%%%%%%%%%%%%%%%%%%%%%%%%%%%%%%%%%%%%%%%

\subsection{Higher-order flow harmonics of prompt photons}

The positions of nucleons in the colliding nuclei are fluctuating. As a consequence there can be higher and/or odd-order flow harmonics in the spectra for each event. For the pure purpose of demonstration and numerical efficiency, here I construct smoothed initial conditions with higher-order symmetries by taking event average with respect to $\Phi_n$ in analogy to the elliptic flow case. The impact parameter is $b = 6$ fm. The spatial anisotropies are $\varepsilon_3 = 0.146$, $\varepsilon_4 = 0.166$ and $\varepsilon_5 = 0.185$ in the aforementioned definition (\ref{eq:en}). Note that the results do not necessarily agree with event-by-event ones quantitatively.

\begin{figure}[tb]
\includegraphics[width=3.0in]{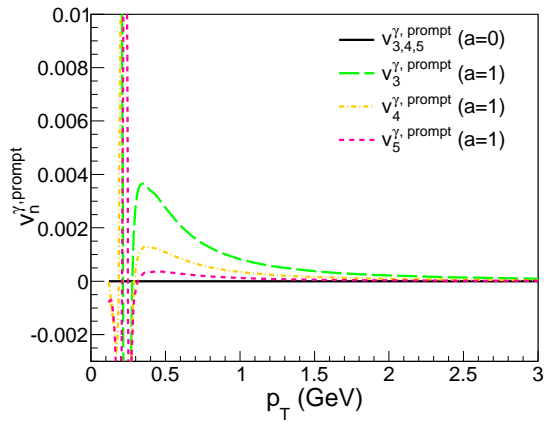}
\caption{(Color online) Triangular flow (dashed line), quadrangular flow (dash-dotted line) and pentagonal flow (dotted line) of prompt photons with medium refraction as a function of the transverse momentum with smoothed initial conditions are compared with those without refraction (solid line). }
\label{fig:vn}
\end{figure}

The triangular flow ($v_3$), the quadrangular flow ($v_4$) and the pentagonal flow ($v_5$) of prompt photons as a function of the transverse momentum for $a = 1$ are shown in Fig.~\ref{fig:vn}. $v_3$ due to the optical effects is positive above $p_T \sim \omega_p$ but again small compared to that of direct photons. $v_4$ and $v_5$ are also found positive in those momentum regions. 
The magnitude of anisotropic flow is found smaller for larger $n$. This behavior can also be found in hadronic flow harmonics. The region near the vanishing momentum, on the other hand, is non-trivial for the higher-order flow harmonics. 

%%%%%%%%%%%%%%%%%%%%%%%%%%%%%%%%%%%%%%%%%%%%%%%%%%%%%%%%%%%%%%%
%%%%%%%%%%%%%%%%%%%%%%%%%%%%%%%%%%%%%%%%%%%%%%%%%%%%%%%%%%%%%%%

%%%%%%%%%%%%%%%%%%%%%%%%%%%%%%%%%%%%%%%%%%%%%%%%%%%%%%%%%%%%%%%%
\section{Discussion and Conclusions}
\label{sec4}
\vspace*{-2mm}
%%%%%%%%%%%%%%%%%%%%%%%%%%%%%%%%%%%%%%%%%%%%%%%%%%%%%%%%%%%%%%%%

Effects of QCD medium refraction on the flow harmonics of direct photons in high-energy heavy-ion collisions are investigated. The mechanism does not modify the total yield of direct photon particle spectra unless the medium becomes absorptive, leaving the hadronic observables unaffected for most of the momentum region.
The numerical estimations with the model refractive index (\ref{eq:index}) imply that the effect of medium refraction for prompt photons can be non-trivial but might not be large enough to explain the direct photon $v_2$ and $v_3$. On the other hand, given that most of the photon $v_2$ enhancement mechanisms so far work only on thermal photons and that prompt photons are considered to have vanishing anisotropy, the prospect of positive prompt photon anisotropy would be important. Also prompt photon $v_n$ does not have to be as large as direct photon $v_n$ because of the large contributions from thermal photons. 

It would be interesting to investigate the ultra-low momentum photon spectra in collider experiments to see whether and/or where $\omega_p$ exists, below which frequency the medium becomes opaque. Heavy-ion collisions at higher energies \cite{Wilde:2012wc} would be able to provide more constraints as $\omega_p$ is implied to be an increasing function of the temperature. This information could be used to constrain the magnitude of refraction effects on direct photon spectra and flow harmonics. Also the photon energy would be disposed in a medium with an opaque or dim region, possibly affecting the bulk medium depending on the characteristic frequency.
It should be stressed that if the actual $\omega_p$ is too small for the current heavy-ion energies %-- though the hard thermal loop calculation implies otherwise -- 
or the QCD refractive index deviates significantly from the present form, the effects would not be observed or appear differently. The refractive index could exhibit, for example, a pseudo-critical behavior as seen in bulk viscosity \cite{Paech:2006st,Monnai:2012jc} and possibly even in photon emission rate \cite{vanHees:2014ida}, which is an interesting prospect.

Further future prospects include application of the method to thermal photons with a more efficient numerical scheme. Also the moving medium would change the phase velocity of light by the drag effect. This is worth-investigating because the relativity is conventionally formulated for tardyons and luxons, and it is not clear if superluminal phase velocity can be na\"{i}vely added to subluminal flow velocity for estimating refraction. 
The effects of refraction by pre- and post-equilibrium media would also be a non-trivial and intriguing issue.

% If you have acknowledgments, this puts in the proper section head.
\begin{acknowledgments}
The work of A.M. is supported by RIKEN Special Postdoctoral Researcher program.
\end{acknowledgments}

% Specify following sections are appendices. Use \appendix* if there
% only one appendix.
%\appendix*

\bibliography{basename of .bib file}

\begin{thebibliography}{99}

\bibitem{RHIC:summary}
%\cite{Adcox:2004mh}
%\bibitem{Adcox:2004mh} 
  K.~Adcox {\it et al.}  [PHENIX Collaboration],
  %``Formation of dense partonic matter in relativistic nucleus-nucleus collisions at RHIC: Experimental evaluation by the PHENIX collaboration,''
  Nucl.\ Phys.\ A {\bf 757}, 184 (2005);
  %[nucl-ex/0410003].
  %%CITATION = NUCL-EX/0410003;%%
%\cite{Adams:2005dq}
%\bibitem{Adams:2005dq} 
  J.~Adams {\it et al.}  [STAR Collaboration],
  %``Experimental and theoretical challenges in the search for the quark gluon plasma: The STAR Collaboration's critical assessment of the evidence from RHIC collisions,''
  Nucl.\ Phys.\ A {\bf 757}, 102 (2005);
  %[nucl-ex/0501009].
  %%CITATION = NUCL-EX/0501009;%%
%\cite{Back:2004je}
%\bibitem{Back:2004je} 
  B.~B.~Back {\it et al.} [PHOBOS Collaboration],
  %``The PHOBOS perspective on discoveries at RHIC,''
  Nucl.\ Phys.\ A {\bf 757}, 28 (2005);
  %[nucl-ex/0410022].
  %%CITATION = NUCL-EX/0410022;%%
  %\cite{Arsene:2004fa}
%\bibitem{Arsene:2004fa} 
  I.~Arsene {\it et al.}  [BRAHMS Collaboration],
  %``Quark gluon plasma and color glass condensate at RHIC? The Perspective from the BRAHMS experiment,''
  Nucl.\ Phys.\ A {\bf 757}, 1 (2005).
  %[nucl-ex/0410020].
  %%CITATION = NUCL-EX/0410020;%%
  
\bibitem{LHC:v2} 
%\cite{Aamodt:2010pa}
%\bibitem{Aamodt:2010pa} 
  K.~Aamodt {\it et al.}  [The ALICE Collaboration],
  %``Elliptic flow of charged particles in Pb-Pb collisions at 2.76 TeV,''
  Phys.\ Rev.\ Lett.\  {\bf 105}, 252302 (2010);
  %[arXiv:1011.3914 [nucl-ex]].
  %%CITATION = ARXIV:1011.3914;%%
%\cite{ATLAS:2011ah}
%\bibitem{ATLAS:2011ah} 
  G.~Aad {\it et al.}  [ATLAS Collaboration],
  %``Measurement of the pseudorapidity and transverse momentum dependence of the elliptic flow of charged particles in lead-lead collisions at $\sqrt{s_{NN}}=2.76$ TeV with the ATLAS detector,''
  Phys.\ Lett.\ B {\bf 707}, 330 (2012);
  %[arXiv:1108.6018 [hep-ex]].
  %%CITATION = ARXIV:1108.6018;%%
%\cite{Chatrchyan:2012wg}
%\bibitem{Chatrchyan:2012wg} 
  S.~Chatrchyan {\it et al.}  [CMS Collaboration],
  %``Centrality dependence of dihadron correlations and azimuthal anisotropy harmonics in PbPb collisions at $\sqrt{s_{NN}}=2.76$ TeV,''
  Eur.\ Phys.\ J.\ C {\bf 72}, 2012 (2012).
  %[arXiv:1201.3158 [nucl-ex]].
  %%CITATION = ARXIV:1201.3158;%%

%\cite{Ollitrault:1992bk}
\bibitem{Ollitrault:1992bk} 
  J.~-Y.~Ollitrault,
  %``Anisotropy as a signature of transverse collective flow,"
  Phys.\ Rev.\ D {\bf 46}, 229 (1992).
  %%CITATION = PHRVA,D46,229;%%
  
%\cite{Poskanzer:1998yz}
\bibitem{Poskanzer:1998yz} 
  A.~M.~Poskanzer and S.~A.~Voloshin,
  %``Methods for analyzing anisotropic flow in relativistic nuclear collisions,"
  Phys.\ Rev.\ C {\bf 58}, 1671 (1998).
  %[nucl-ex/9805001].
  %%CITATION = NUCL-EX/9805001;%%

%\cite{Kolb:2000fha}
\bibitem{Kolb:2000fha} 
  P.~F.~Kolb, P.~Huovinen, U.~W.~Heinz and H.~Heiselberg,
  %``Elliptic flow at SPS and RHIC: From kinetic transport to hydrodynamics,''
  Phys.\ Lett.\ B {\bf 500}, 232 (2001);
  %[hep-ph/0012137].
  %%CITATION = HEP-PH/0012137;%%
%
%\cite{Schenke:2010rr}
%\bibitem{Schenke:2010rr} 
  B.~Schenke, S.~Jeon and C.~Gale,
  %``Elliptic and triangular flow in event-by-event (3+1)D viscous hydrodynamics,''
  Phys.\ Rev.\ Lett.\  {\bf 106}, 042301 (2011).
  %[arXiv:1009.3244 [hep-ph]].
  %%CITATION = ARXIV:1009.3244;%%
  
%\cite{Adare:2011zr}
\bibitem{Adare:2011zr} 
  A.~Adare {\it et al.}  [PHENIX Collaboration],
  %``Observation of direct-photon collective flow in $\sqrt{s_{NN}}=200$ GeV Au+Au collisions,''
  Phys.\ Rev.\ Lett.\  {\bf 109}, 122302 (2012).
  %[arXiv:1105.4126 [nucl-ex]].
  %%CITATION = ARXIV:1105.4126;%%
  
%\cite{Lohner:2012ct}
\bibitem{Lohner:2012ct} 
  D.~Lohner [ALICE Collaboration],
  %``Measurement of Direct-Photon Elliptic Flow in Pb-Pb Collisions at $\sqrt{s_{NN}} = 2.76$ TeV,''
  J.\ Phys.\ Conf.\ Ser.\  {\bf 446}, 012028 (2013).
  %[arXiv:1212.3995 [hep-ex]].
  %%CITATION = ARXIV:1212.3995;%%
  
%\cite{Chatterjee:2005de}
\bibitem{Chatterjee:2005de} 
  R.~Chatterjee, E.~S.~Frodermann, U.~W.~Heinz and D.~K.~Srivastava,
  %``Elliptic flow of thermal photons in relativistic nuclear collisions,''
  Phys.\ Rev.\ Lett.\  {\bf 96}, 202302 (2006).
  %[nucl-th/0511079].
  %%CITATION = NUCL-TH/0511079;%%
  
%\cite{Chatterjee:2008tp}
\bibitem{Chatterjee:2008tp} 
  R.~Chatterjee and D.~K.~Srivastava,
  %``Elliptic flow of thermal photons and formation time of quark gluon plasma at RHIC,''
  Phys.\ Rev.\ C {\bf 79}, 021901 (2009).
  %[arXiv:0809.0548 [nucl-th]].
  %%CITATION = ARXIV:0809.0548;%%
  
%\cite{Chatterjee:2011dw}
\bibitem{Chatterjee:2011dw} 
  R.~Chatterjee, H.~Holopainen, T.~Renk and K.~J.~Eskola,
  %``Enhancement of thermal photon production in event-by-event hydrodynamics,''
  Phys.\ Rev.\ C {\bf 83}, 054908 (2011).
  %[arXiv:1102.4706 [hep-ph]].
  %%CITATION = ARXIV:1102.4706;%%
  
%\cite{Holopainen:2011pd}
\bibitem{Holopainen:2011pd} 
  H.~Holopainen, S.~Rasanen and K.~J.~Eskola,
  %``Elliptic flow of thermal photons in heavy-ion collisions at Relativistic Heavy Ion Collider and Large Hadron Collider,''
  Phys.\ Rev.\ C {\bf 84}, 064903 (2011).
  %[arXiv:1104.5371 [hep-ph]].
  %%CITATION = ARXIV:1104.5371;%%
  
%\cite{Chatterjee:2013naa}
\bibitem{Chatterjee:2013naa} 
  R.~Chatterjee, H.~Holopainen, I.~Helenius, T.~Renk and K.~J.~Eskola,
  %``Elliptic flow of thermal photons from event-by-event hydrodynamic model,''
  Phys.\ Rev.\ C {\bf 88}, 034901 (2013).
  %[arXiv:1305.6443 [hep-ph]].
  %%CITATION = ARXIV:1305.6443;%%
  
%\cite{Chatterjee:2014nta}
\bibitem{Chatterjee:2014nta} 
  R.~Chatterjee, D.~K.~Srivastava and T.~Renk,
  %``Triangular flow of thermal photons from an event-by-event hydrodynamic model for 2.76A TeV Pb+Pb collisions at LHC,''
  arXiv:1401.7464 [hep-ph].
  
%\cite{Liu:2009kta}
\bibitem{Liu:2009kta} 
  F.~-M.~Liu, T.~Hirano, K.~Werner and Y.~Zhu,
  %``Elliptic flow of thermal photons in Au + Au collisions at s(NN)**(1/2) = 200-GeV,''
  Phys.\ Rev.\ C {\bf 80}, 034905 (2009).
  %[arXiv:0902.1303 [hep-ph]].
  %%CITATION = ARXIV:0902.1303;%%
  
%\cite{vanHees:2011vb}
\bibitem{vanHees:2011vb} 
  H.~van Hees, C.~Gale and R.~Rapp,
  %``Thermal Photons and Collective Flow at the Relativistic Heavy-Ion Collider,''
  Phys.\ Rev.\ C {\bf 84}, 054906 (2011).
  %[arXiv:1108.2131 [hep-ph]].
  %%CITATION = ARXIV:1108.2131;%%
  
%\cite{Dion:2011pp}
\bibitem{Dion:2011pp} 
  M.~Dion, J.~-F.~Paquet, B.~Schenke, C.~Young, S.~Jeon and C.~Gale,
  %``Viscous photons in relativistic heavy ion collisions,''
  Phys.\ Rev.\ C {\bf 84}, 064901 (2011).
  %[arXiv:1109.4405 [hep-ph]].
  %%CITATION = ARXIV:1109.4405;%%
  
%\cite{Basar:2012bp}
\bibitem{Basar:2012bp} 
  G.~Basar, D.~Kharzeev, and V.~Skokov,
  %``Conformal anomaly as a source of soft photons in heavy ion collisions,''
  Phys.\ Rev.\ Lett.\  {\bf 109}, 202303 (2012);
  %[arXiv:1206.1334 [hep-ph]].
  %%CITATION = ARXIV:1206.1334;%%
%
%\cite{Basar:2014swa}
%\bibitem{Basar:2014swa} 
  G.~Basar, D.~E.~Kharzeev and E.~V.~Shuryak,
  %``The Magneto-Sono-Luminescence and its signatures in photon and dilepton production in heavy ion collisions,''
  Phys.\ Rev.\ C {\bf 90}, 014905 (2014).
  %[arXiv:1402.2286 [hep-ph]].
  %%CITATION = ARXIV:1402.2286;%%
  
%\cite{Bzdak:2012fr}
\bibitem{Bzdak:2012fr} 
  A.~Bzdak and V.~Skokov,
  %``Anisotropy of photon production: initial eccentricity or magnetic field,''
  Phys.\ Rev.\ Lett.\  {\bf 110}, 192301 (2013).
  %[arXiv:1208.5502 [hep-ph]].
  %%CITATION = ARXIV:1208.5502;%%
  
%\cite{Goloviznin:2012dy}
\bibitem{Goloviznin:2012dy} 
  V.~V.~Goloviznin, A.~M.~Snigirev and G.~M.~Zinovjev,
  %``Towards azimuthal anisotropy of direct photons,''
  JETP Lett.\  {\bf 98}, 61 (2013).
  %[arXiv:1209.2380 [hep-ph]].
  %%CITATION = ARXIV:1209.2380;%%
  
%\cite{Hattori:2012je}
\bibitem{Hattori:2012je} 
  K.~Hattori and K.~Itakura,
  %``Vacuum birefringence in strong magnetic fields: (I) Photon polarization tensor with all the Landau levels,''
  Annals Phys.\  {\bf 330}, 23 (2013);
  %[arXiv:1209.2663 [hep-ph]].
  %%CITATION = ARXIV:1209.2663;%%
  %
%\cite{Hattori:2012ny}
%\bibitem{Hattori:2012ny} 
  %K.~Hattori and K.~Itakura,
  %``Vacuum birefringence in strong magnetic fields: (II) Complex refractive index from the lowest Landau level,''
  Annals Phys.\  {\bf 334}, 58 (2013).
  %[arXiv:1212.1897 [hep-ph]].
  %%CITATION = ARXIV:1212.1897;%%
  
%\cite{Liu:2012ax}
\bibitem{Liu:2012ax} 
  F.~-M.~Liu and S.~-X.~Liu,
  %``Study the early stage of heavy ion collisions with direct photons,''
  Phys.\ Rev.\ C {\bf 89}, 034906 (2014).
  %[arXiv:1212.6587 [nucl-th]].
  %%CITATION = ARXIV:1212.6587;%%
  
  %\cite{Shen:2013cca}
\bibitem{Shen:2013cca} 
  C.~Shen, U.~W.~Heinz, J.~-F.~Paquet, I.~Kozlov and C.~Gale,
  %``Anisotropic flow of thermal photons as a quark-gluon plasma viscometer,''
  arXiv:1308.2111 [nucl-th].
  %%CITATION = ARXIV:1308.2111;%%
  
%\cite{Muller:2013ila}
\bibitem{Muller:2013ila} 
  B.~M\"{u}ller, S.~-Y.~Wu and D.~-L.~Yang,
  %``Elliptic flow from thermal photons with magnetic field in holography,''
  Phys.\ Rev.\ D {\bf 89}, 026013 (2014).
  %[arXiv:1308.6568 [hep-th]].
  %%CITATION = ARXIV:1308.6568;%%
  
%\cite{Linnyk:2013hta}
\bibitem{Linnyk:2013hta} 
  O.~Linnyk, V.~P.~Konchakovski, W.~Cassing and E.~L.~Bratkovskaya,
  %``Photon elliptic flow in relativistic heavy-ion collisions: hadronic versus partonic sources,''
  Phys.\ Rev.\ C {\bf 88}, 034904 (2013);
  %[arXiv:1304.7030 [nucl-th]].
  %%CITATION = ARXIV:1304.7030;%%
%
%\cite{Linnyk:2013wma}
%\bibitem{Linnyk:2013wma} 
  O.~Linnyk, W.~Cassing and E.~Bratkovskaya,
  %``Centrality dependence of the direct photon yield and elliptic flow in heavy-ion collisions at sqrt(s)=200 GeV,''
  arXiv:1311.0279 [nucl-th].
  %%CITATION = ARXIV:1311.0279;%%
  
%\cite{Monnai:2014kqa}
\bibitem{Monnai:2014kqa} 
  A.~Monnai,
  %``Thermal photon $v_2$ with slow quark chemical equilibration,''
  arXiv:1403.4225 [nucl-th].
  %%CITATION = ARXIV:1403.4225;%%
  
%\cite{McLerran:2014hza}
\bibitem{McLerran:2014hza} 
  L.~McLerran and B.~Schenke,
  %``The Glasma, Photons and the Implications of Anisotropy,''
  arXiv:1403.7462 [hep-ph].
  %%CITATION = ARXIV:1403.7462;%%
  
%\cite{vanHees:2014ida}
\bibitem{vanHees:2014ida} 
  H.~van Hees, M.~He and R.~Rapp,
  %``Pseudo-Critical Enhancement of Thermal Photons in Relativistic Heavy-Ion Collisions,''
  arXiv:1404.2846 [nucl-th].
  
%\cite{Klimov:1982bv}
\bibitem{Klimov:1982bv} 
  V.~V.~Klimov,
  %``Collective Excitations in a Hot Quark Gluon Plasma,''
  Sov.\ Phys.\ JETP {\bf 55}, 199 (1982).
  %[Zh.\ Eksp.\ Teor.\ Fiz.\  {\bf 82}, 336 (1982)].
  
%\cite{Liu:2011if}
\bibitem{Liu:2011if} 
  J.~Liu, M.~-j.~Luo, Q.~Wang and H.~-j.~Xu,
  %``Refractive Index of Light in the Quark-Gluon Plasma with the Hard-Thermal-Loop Perturbation Theory,''
  Phys.\ Rev.\ D {\bf 84}, 125027 (2011).
  %[arXiv:1109.4083 [hep-ph]].
  %%CITATION = ARXIV:1109.4083;%%
  
%\cite{Jiang:2013nw}
\bibitem{Jiang:2013nw} 
  B.~-f.~Jiang, D.~-f.~Hou, J.~-r.~Li and Y.~-J.~Gao,
  %``Refractive index in the viscous quark-gluon plasma,''
  Phys.\ Rev.\ D {\bf 88}, 045014 (2013).
  %[arXiv:1301.1789 [hep-ph]].
  %%CITATION = ARXIV:1301.1789;%%
  
%\cite{Adare:2008ab}
\bibitem{Adare:2008ab} 
  A.~Adare {\it et al.}  [PHENIX Collaboration],
  %``Enhanced production of direct photons in Au+Au collisions at $\sqrt{s_{NN}}=200$ GeV and implications for the initial temperature,''
  Phys.\ Rev.\ Lett.\  {\bf 104}, 132301 (2010).
  %[arXiv:0804.4168 [nucl-ex]].
  %%CITATION = ARXIV:0804.4168;%%
  
%\cite{Adare:2014fwh}
\bibitem{Adare:2014fwh} 
  A.~Adare {\it et al.}  [PHENIX Collaboration],
  %``Centrality dependence of low-momentum direct-photon production in Au$+$Au collisions at $\sqrt{s_{_{NN}}}=200$ GeV,''
  arXiv:1405.3940 [nucl-ex].
  %%CITATION = ARXIV:1405.3940;%%
  
%\cite{Turbide:2003si}
\bibitem{Turbide:2003si} 
  S.~Turbide, R.~Rapp and C.~Gale,
  %``Hadronic production of thermal photons,''
  Phys.\ Rev.\ C {\bf 69}, 014903 (2004);
  %[hep-ph/0308085].
  %%CITATION = HEP-PH/0308085;%%
%  

%\cite{Yao:2006px}
\bibitem{Yao:2006px} 
  W.~M.~Yao {\it et al.}  [Particle Data Group Collaboration],
  %``Review of Particle Physics,''
  J.\ Phys.\ G {\bf 33}, 1 (2006).
  %%CITATION = JPHGB,G33,1;%%

%\cite{Borsanyi:2010cj}
\bibitem{Borsanyi:2010cj} 
  S.~Borsanyi, G.~Endrodi, Z.~Fodor, A.~Jakovac, S.~D.~Katz, S.~Krieg, C.~Ratti and K.~K.~Szabo,
  %``The QCD equation of state with dynamical quarks,''
  JHEP {\bf 1011}, 077 (2010).
  %[arXiv:1007.2580 [hep-lat]].
  %%CITATION = ARXIV:1007.2580;%%

%\cite{Miller:2007ri}
\bibitem{Miller:2007ri} 
  M.~L.~Miller, K.~Reygers, S.~J.~Sanders and P.~Steinberg,
  %``Glauber modeling in high energy nuclear collisions,''
  Ann.\ Rev.\ Nucl.\ Part.\ Sci.\  {\bf 57}, 205 (2007).
  %[nucl-ex/0701025].
  %%CITATION = NUCL-EX/0701025;%%
  
%\cite{Teaney:2010vd}
\bibitem{Teaney:2010vd} 
  D.~Teaney and L.~Yan,
  %``Triangularity and Dipole Asymmetry in Heavy Ion Collisions,''
  Phys.\ Rev.\ C {\bf 83}, 064904 (2011).
  %[arXiv:1010.1876 [nucl-th]].
  %%CITATION = ARXIV:1010.1876;%%
  
%\cite{Wilde:2012wc}
\bibitem{Wilde:2012wc} 
  M.~Wilde [ALICE Collaboration],
  %``Measurement of Direct Photons in pp and Pb-Pb Collisions with ALICE,''
  Nucl.\ Phys.\ A {\bf 904-905}, 573c (2013).
  %[arXiv:1210.5958 [hep-ex]].
  %%CITATION = ARXIV:1210.5958;%%

\bibitem{Paech:2006st}
  K.~Paech and S.~Pratt,
  %``Origins of bulk viscosity at RHIC,"
  Phys.\ Rev.\  C {\bf 74}, 014901 (2006).
  %[arXiv:nucl-th/0604008].
  %%CITATION = PHRVA,C74,014901;%%

%\cite{Monnai:2012jc}
\bibitem{Monnai:2012jc} 
  A.~Monnai,
  %``Dissipative Hydrodynamic Effects on Baryon Stopping,''
  Phys.\ Rev.\ C {\bf 86}, 014908 (2012).
  %[arXiv:1204.4713 [nucl-th]].
  %%CITATION = ARXIV:1204.4713;%%
  
\end{thebibliography}

\end{document}